 \documentstyle[11pt,aaspp4]{article}   

\slugcomment{ApJ, in press, vol 512, 1999}

\lefthead{Fenimore et al.}
\righthead{GRB Surface Filling Factor}
\begin{document}

\title{Gamma-Ray Bursts and Relativistic Shells:\\
The Surface Filling Factor}
\author{E.~E.~Fenimore, C. Cooper, E.~Ramirez-Ruiz, and M.~C.~Sumner}
\affil{MS D436, Los Alamos National Laboratory, Los Alamos, NM 87545}

\and

\author{A. Yoshida and M. Namiki}
\affil{The Institute of Physical and Chemical Research (RIKEN) \\
2-1, Hirosawa, Wako, Saitama 351-0198, Japan}

\begin{abstract}
The variability observed in many complex gamma-ray bursts (GRBs) is
inconsistent with causally connected variations in  a single,
symmetric, relativistic shell interacting with the ambient material
(``external shocks''). Rather, either the central site must
produce $\sim 10^{50}$ erg s$^{-1}$ for hundreds of seconds
(``internal shocks'') or the
local spherical symmetry of the shell must be 
broken on an angular scale much smaller
than $\Gamma^{-1}$ where $\Gamma$ is the bulk Lorentz factor for the
shell. The observed variability in the external shock models 
arises from the number of causally
connected regions that (randomly) become active.  We define the
``surface filling factor'' to be the ratio of the area of causally 
connected regions that become active to the observable area of the
shell. From the observed variability in 52 BATSE bursts, we estimate
the surface filling factor to be typically $\sim 5 \times 10^{-3}$ although
some values are near unity.  We find that the surface filling factor, $f$, is
$\sim 0.1 \Delta T/T$ in both the constant $\Gamma$ phase (which
probably produces the GRB) and the
deaccelerating phase (which probably produces the x-ray afterglows).
Here, $\Delta T$ is a typical time scale of variability and $T$ is the
time since the initial signal. We analyze the 2 hr flare seen by ASCA
36 hr after the GRB and conclude that the surface filling factor must be small
($10^{-3}$) in
the x-ray afterglow phase as well.
Compared to the energy required for an isotropic shell, $E_{\rm iso}$,
explanations for low surface
filling factor
can either require more energy 
($f^{-1}E_{\rm iso} \sim 10^{56}$ erg) or less energy
($({\Delta T \over 4T})^2E_{\rm iso} \sim 10^{49}$ erg).
Thus, the low filling factor cannot be used as a strong argument that GRBs must
be internal shocks.
\end{abstract}

\keywords{gamma-rays: bursts}

\newcount\eqnumber
\eqnumber=1
\def\neweq{{\the\eqnumber}\global\advance\eqnumber by 1}
\def\eqnam#1#2{\xdef#1{\the\eqnumber}}
\def\lasteq{\advance\eqnumber by -1 {\the\eqnumber}\advance
    \eqnumber by 1}
\def\domega{{\rm d}$\Omega$}
\def\Mesz{M\'esz\'aros}
\def\INITIALCON{{4\pi E_{54} \over \rho_0{\rm d}\Omega}}
\section{INTRODUCTION}

Gamma-ray burst (GRB) spectra often extend to very 
high energies with no indication of
attenuation by photon-photon interactions.
This implies substantial relativistic
bulk motion of the radiating material with Lorentz factors
in the range of
$10^2$ to $10^3$.
At cosmological distances, GRBs require an energy reservoir on the order of
$10^{52}$ erg.  The likely sources of such a reservoir would be the rest
mass of a compact object released during a merger 
(e.g., either neutron star -- neutron star or neutron star -- black hole).
However, most detailed calculations of mergers occur on time 
scales of less than 1 s, much less than the observed durations of GRBs
(often $10^2$ s and sometimes $10^3$ s).

Two classes of models have arisen that explain various (but not all)
aspects of the observations.
In the ``external'' shock models (\cite{mr93}),
the release of energy during the merger is very quick, and a relativistic
shell forms that expands outward for a long period
of time ($10^5$ to $10^7$
s).
At some point, interactions with the external medium
(hence the name)
cause the energy of the bulk motion to be converted to gamma-rays.
Although the shell might produce gamma-rays for a long period of time,
the shell keeps up with the photons
such that they arrive at a detector over a relatively short period of time.
If the shell has a velocity, $v = \beta c$, with a corresponding 
bulk Lorentz factor,
$\Gamma = (1-\beta^2)^{-1/2}$,  then photons emitted over a period $t$
arrive at a detector over a much shorter period, $T = (1-\beta)t = 
t/(2\Gamma^2)$.
Although this model is consistent with the short energy release expected
for a merger and the observed long time scale of 
GRBs, we have argued that it cannot
explain the long complex time histories except under extreme conditions
(\cite{fmn96}).  In particular, we argue from
kinematic considerations that the long
gaps often seen in GRBs are inconsistent with local spherical
symmetry,
that the rapid time variability implies that only a small fraction of the
shell becomes active, and that the
observed average envelope of emission is
inconsistent with that expected from a relativistic shell.  These arguments are
contained in \cite{fmn96n,fs97n},
and \cite{frs98n}.
In addition, \cite{dar97n}, \cite{sp97bn}, and \cite{kps97n}  
argue from hydrodynamic
considerations that the external shock model cannot provide the
energy or timescales observed in GRBs.

The alternative theory is that a central site releases energy in the form
of a wind or multiple shells over a period of time commensurate with the
observed duration of the GRB (\cite{rm94}).
The gamma-rays are produced by the 
internal interactions within the wind; hence these scenarios are often 
referred to as internal shock models.  These models have two weaknesses:
first, there is a concern that internal shocks are rather inefficient
(although, see \cite{kps97}), and
second, the long, complex time history of a GRB  must be postulated
at the central site.
On the other hand, the free parameters associated with the multiple shells
can probably explain any observed time history (\cite{kps97}).

The need to explain long time histories was further complicated by the
discovery of the x-ray afterglows lasting hours (\cite{cos97}),
the optical afterglows lasting weeks to months (\cite{metzger97}),
and the radio
afterglows lasting many months (\cite{frail97}).
These extended time scales
appear too long to have been produced by a lingering central site made by a
merger.
In addition, the observed power law decay is expected from many
external shock models
(\cite{wrm97,danr97,tavani97,wkf98,pm98,mrw98,rm98,spn98}).
\cite{ps97n} suggested
that the initial gamma-ray phase is due to internal shocks from
a relativistic wind (or multiple
shells) that merge into a single relativistic shell which
then produces the afterglows
in a manner similar to the external shock models.
This model avoids the difficulty
of explaining the rapid variability of the gamma-ray phase with a
single shell while retaining the long time scale capability of the
single shell for the afterglows.

The purpose of this paper is to demonstrate that the
external shock model cannot utilize the full surface of shells.
Fundamentally, it arises because of the relatively rapid time
variability of the gamma-ray phase. \cite{sp97a} use this as an argument 
that external shock models would require too much energy.
However, this should not be considered a strong argument that the GRB 
phase is not external shocks since there are several scenarios that can 
utilize only a portion of the shell's surface.
 We also analyze the time variability recently observed
by ASCA in the x-ray afterglow of GB970828 (\cite{yoshida98}) and show that
its time variability implies low surface utilization  in the
x-ray afterglow phase as well.

\section{RAPID VARIABILITY FROM RELATIVISTIC SHELLS}

To understand the problem caused by rapid variability, one must
emphasize the difference between arrival time at the detector 
(which we denote with
$T$) and coordinate time (equivalent to laboratory time,
the detector rest frame time, {\it and} the rest frame of the
central explosion,  denoted by $t$).
Coordinate time is measured by clocks placed at all locations within
the frame and can measure when the photons were produced.  
In contrast, a detector (such as BATSE) measures when the photons
arrive at a single location.
These two times are related as $T = (1-\beta\cos\theta)t$
where $\theta$ is the angle between the direction of motion of the
emitting material and the direction to the detector.
The large bulk Lorentz
factor means that the shell is almost certainly moving directly at the
observer's detector (i.e., we must be looking at the shell head-on
rather than looking at the sides of a jet, so $\theta$ is small.
When $\theta$ is 0, $T = t/(2\Gamma^2)$.  In contrast, the Lorentz
transformation between the time in the rest frame of the detector ($t$) and
time in a frame moving with the shell ($t^\prime$), is $t = \Gamma t^\prime$.
Consider the early behavior of the shell.
Assume that clocks are placed along the path of the shell in
the detector's rest frame.  These clocks report coordinate time, that
is, the time the shell passes the location of the clocks.  
Consider an extreme case such as $\Gamma = 10^3$.  Clocks placed with a
spacing of
$3 \times 10^{16}$ cm will see the shell pass their locations at times
separated by
$10^6$ s.  Because the shell is viewed by the detector head-on, its
movement causes the shell to keep up with the photons it produces such
that photons that were separated by $10^6$ in the detector's rest frame,
arrive at the detector separated by
a  much shorter interval  $T = (1-\beta)t
= t/(2\Gamma^2)$.  If the shell expands in the detector rest frame for
$10^7$ seconds, the photons all arrive at the detector within 5 s.
Note that this contraction of the photon arrival time is not
the result of a Lorentz transformation. Rather,
it reflects that a detector at a single location measures when photons
arrive, not necessarily the relative time they were emitted. 

In the early phase of the shell's expansion, $\Gamma$ is effectively
constant ($=\Gamma_0$) such that $t=T/(2\Gamma^2_0)$.  Eventually, the
shell begins to deaccelerate as it sweeps up the interstellar medium (ISM).
Without detailed knowledge of the physical process that generates the 
gamma rays, it is not clear if the expansion is adiabatic or dominated
by radiation losses.  However, both solutions lead to a power law
decay (\cite{spn98,mrw98,rm98}), and we will 
show that our conclusions only depend on
the power law nature of the decay and not the exact index of the power
law. For the sake of an example, we will
assume that the expansion is adiabatic. The energy required to sweep up an ISM
mass of $m_s$ is $\Gamma^2 m_s c^2$ (\cite{mr97}).
Let $E_0$ be the initial energy
deposited into the shell such that $E_0 = \Gamma_0 M_s c^2$, where
$M_s$ is the original mass of the shell. If
 $\rho$ is the density of the ISM, one can solve for the radius of
the shell as a function of the coordinate time by solving 
\eqnam{\ECONSERVE}{ECONSERVE}
$$
E_0 = \Gamma(t) M_s c^2 +\Gamma^2(t) \rho 
{{\rm d}\Omega \over 3}r^3(t)c^2 \eqno(\neweq)
$$
Here, \domega~is the angular size of the ejecta to account for the
situation where the expansion is mechanically confined to a jet.
The arrival time of the photons as a function of the coordinate time
that the photons were emitted can be found from
\eqnam{\DTDT}{DTDT}
$$
{dT \over dt} = 1-\beta(t) = {1 \over 2\Gamma^2(t)} \eqno(\neweq)
$$
and ${dr \over dt} = c\beta(t)$.
Figure 1 shows $t$ as a function of $T$ for a typical situation 
($4\pi E_0/{\rm d}\Omega = 10^{54}$ erg, $\rho = 1$
atom cm$^{-3}$).
It is unclear what fraction of the shell's energy is converted to gamma-rays.
Observations imply that the energy in gamma-rays {\it if the explosion
is isotropic} is on the order of $10^{52}$ erg.  If \domega~ is much less
than $4\pi$, much less energy is required (implying many more unseen
events).  The observations only limit the energy per $4\pi$ steradian
to be $\sim 10^{52}$ erg. We
scale our solutions to $4\pi E_0/{\rm d}\Omega = 10^{54}$ erg
which assumes that $\sim$ 1\% of the energy is
converted to gamma-rays.

 There are three phases of the $t-T$ relationship:
the initial constant-$\Gamma$ phase, a phase where
$t = 1.63 \times 10^6 (\INITIALCON)^{1/4} T^{1/4}$, and
a final phase when the
shell becomes nonrelativistic where $t$ is $T$ plus a constant.
Here, $E_{54}$ is $E_0$ in units of $10^{54}$ erg and $\rho_0$ is the
ISM density in units of 1 atom cm$^{-3}$.
The second and third phase is only weakly dependent on the initial
conditions. Neither depends on the initial $\Gamma_0$, and they depend on
the other parameters to the 1/4 power.  Thus, except for the duration
of the first phase, different bursts ought to have similar $t-T$
relationships.  If one assumes the extreme nonadiabatic case that the
momentum of the shell is conserved rather than the energy, then the
1/4 power becomes 1/7. Many other indexes as possible (see 
\cite{spn98,mrw98,rm98}).

Although the exact
nature of the conversion of bulk motion to gamma-rays is unclear, 
we  argue that the gamma-ray phase must be contained in the first phase
where $t =
T/(2\Gamma_0^2)$. We note that many GRBs show a similar 
time of arrival structure
(e.g., observed subpeak widths) at
the end of the burst and at the beginning of the burst.
In the first phase, the observed arrival time 
structure ($\Delta T$) is related to the
emission time as $\Delta t/(2\Gamma_0^2)$ such that a constant subpeak
width can occur if the emission time is constant in coordinate time.
In the second phase, $\Delta T \propto t^3\Delta t$ such that constant
subpeak width can only occur if the emission time
($\Delta t$) scales as $t^{-3}$.  This is consistent with our scaling
of $4\pi E_0/{\rm d}\Omega = 10^{54}$ erg, which allows the first phase
to last 200 s (see Figure 1).
The afterglows probably occur in the
$t \propto T^{1/4}$ phase where $\Gamma$ is more modest.

As one can see from Figure 1, our view of GRBs becomes quite distorted
because we only measure arrival time at the earth rather than the time
of emission in the earth's rest frame (remember, 
no Lorentz transformations are involved).  It is often said that GRBs are
quite quick, say 30 s.  Actually, in the external shock models,
they occur longer by $2\Gamma^2$
or, perhaps, up to $6 \times 10^7$ s if $\Gamma$ is $10^3$.  The
afterglow period seems to cover a very large dynamic range from $10^3$
s to $10^7$ s.  Actually, because $t \propto T^{1/4}$, the dynamic
range is only a factor of 10.  

\section{RELATIONSHIP BETWEEN THE SURFACE FILLING\\
 FACTOR AND THE OBSERVATIONS}

In the external shock model, there is a single shell, and the rapid
variability we observe can only be accounted for by a 
nonuniform production of
gamma-rays by the shell.  Consider BATSE bursts 1883 and 2993 shown in
Figure 2. They have very similar overall envelopes of
emission and similar intensities.  Burst 2993, however, has stronger
deviations from the envelope in the form of multiple peaks. Presumably
this structure arises because various patches (or emitting
``entities'') on the shell randomly become active.  In the case of
burst 1883, so many entities become simultaneously active that the
overall envelope appears quite smooth, whereas in the case of burst 2993, fewer
entities become active so random fluctuations in the number of 
simultaneously active entities cause the peak structure to be spiky.  
In this view, 
each observed peak in not necessarily caused by a single entity, but the
peak structure is caused by random variations in the number of active
entities (see \cite{sf98}).

In \cite{fmn96n} and \cite{fs97n},
we showed that the expected
envelope  of the GRB phase has  a ``FRED'' like shape (Fast Rise, 
Exponential 
Decay although the decays are often just slower, not necessarily
exponential). The expected envelope
is a power law, $T^{-\alpha-1}$, where $\alpha$ is the spectral
number index (typically $\sim 1.5$).  (Note that this power law refers to 
the GRB phase in constrast to the well observed power law decay found in 
the afterglows.) In \cite{sf98n},
we showed that the volume that contributes to a set period of time,
$\Delta T$, is constant.  The $T^{-\alpha -1}$ envelope
comes from the beaming.  Thus, if the entities form randomly, one
can expect that the rate of entities causing peaks is effectively
constant during the decay phase.  Indeed, the presence of long gaps
in GRBs is one of our key arguments that the GRB phase is not due to a 
simple single shell. Gaps require large causally disconnected regions to
coordinate their lack of emission (see \cite{fmn96}
and \cite{frs98}).

We define the ``surface filling factor'' as the fraction of the
shell's surface
that becomes active.  Let $A_N$ be the area of an entity and $N_N$ be
the number of entities that (randomly) become active during the interval
$T_{\rm obs}$.  If $A_{\rm obs}$ is the area of the shell that can contribute
during $T_{\rm obs}$, then the surface filling factor is
\eqnam{\EFF}{EFF}
$$
f = N_N{A_N \over A_{\rm obs}}~= ~~N_N{A_N \over \eta A_S}~
\eqno(\neweq)
$$
where $\eta$ is the fraction of the visible area of the shell ($A_S$) that
contributes during the interval $T_{\rm obs}$.

The number of entities can be determined from the observed fluctuations
in the time history.  The overall envelope is due to the beaming, so
the residue variations about the overall envelope are due to a
combination of the Poisson variations in the number of entities and
the Poisson variations associated with the count statistics.
Assume for the moment that the contribution due to the count statistics
are small such that the variations come from the Poisson variations in
the number of contributing entities.
We first remove the envelope by fitting a polynomial function to it.
The observations are divided by the polynomial function,
so the result is a flat
envelope with variations due to the number of entities that, on
average, are active simultaneously.  The rate of occurrence of the
entities, $\mu_N$, can be found directly from the variations because the
observed mean level is $K\mu_N$ where $K$ is some constant and the
variance is $K^2\mu_N$. We define $N$ to be the observed mean level of
the flattened envelope
and $\delta N$ to be the root mean square of the flattened envelope.
In the case of no contribution from the counting statistics, 
$\mu_N$ would be  $(N/\delta N)^2$.
Here, we have implicitly assumed that all entities are identical.
This is supported
by the fact that peaks within GRBs usually are similar to each other.
The actual root mean square of the flattened envelope is a combination
of the variance due to the entities and the variance due to the counting
statistics, $\sigma^2_{CS}$. We assume they add in quadrature, that is, 
$N^2/(\delta N)^2 = \mu_N + \sigma_{CS}^2$.  
We estimate $\sigma_{CS}$ to be root mean square of many Monte Carlo
realizations of the flattened envelope.
The rate of occurrence of entities is
\eqnam{\ENTITYRATE}{ENTITYRATE}
$$
\mu_N = {N^2 \over (\delta N)^2} - \sigma_{CS}^2~~.
\eqno(\neweq)
$$
This rate is the number of events per the time scale of the entities.  
Thus, the total number of 
entities that occur
within a period $T_{\rm obs}$ is
\eqnam{\ENTITYNUM}{ENTITYNUM}
$$
N_N = \mu_N{T_{\rm obs} \over \Delta T_p}
\eqno(\neweq)
$$
where $\Delta T_p$ is the time scale for a single entity.

We have
consider several scenarios that relate the size of an entity causing
a peak to an observed peak duration, $\Delta T_p$ (see Table 1 in
\cite{fmn96}). Here, we consider the two most
likely scenarios for the formation of a peak:  regions that grow and 
regions formed by the interaction with the ISM.

Consider a gamma-ray producing region that grows at the speed of
sound, $c_s$, for a period $\Delta t^\prime$ in the rest frame of the
shell where presumably the region is symmetric.  This growth might be
associated with a developing shock.  Let $\Delta
r^\prime_\parallel$ be the radius of the region in the rest frame along the
direction of the motion and $\Delta r^\prime_\perp$ be the radius in the
perpendicular direction such that
\eqnam{\LORENTZ}{LORENTZ}
$$
\Delta r^\prime_\parallel = \Delta r^\prime_\perp = c_s\Delta t^\prime
= c_s \Gamma^{-1} \Delta t~~.
\eqno(\neweq)
$$
The sizes in the rest frame of the detector are related to the sizes in
the rest frame of the shell as: $\Delta r_\parallel =
\Gamma^{-1}\Delta r^\prime_\parallel$ and 
$\Delta r_\perp = \Delta r^\prime_\perp$.

Combining the effects of the movement of the shell during the growth
with the maximum size that the entity can grow in time $\Delta t$, we
find that the duration in arrival time is (see Table 1 of \cite{fmn96}):
\eqnam{\DTGROWTH}{DTGROWTH}
$$
\Delta T_p = {\Delta t \over \Gamma^2}
\bigg[\big({1 \over 2}\big)^2 + \big({c_s \over c}\big)^2\bigg]^{1/2}~.
\eqno(\neweq)
$$ 
The size of the emitting entity is
\eqnam{\SIZEA}{SIZEA}
$$
A_N = \pi \Delta r^2_\perp = \pi
\big({c_s \Delta t \over \Gamma}\big)^2 =
{\pi c_s^2 \Gamma^2 \Delta T_p^2 \over
\big[({1 \over 2})^2 + ({c_s \over c})^2\big]}~~.
\eqno(\neweq)
$$

The alternative cause of a peak resulting from the shell is that the shell
interacts with an ambient object such as an ISM cloud.  Presumably, the
object is symmetric such that $\Delta R_{\perp} = \Delta R_{\parallel}
= \Delta R_{\rm amb}$. (We use lower case $\Delta r$ for an object
that grows in a shell and upper case $\Delta R$ for an ambient
object.)
 We assume the case of a ``collapsible'' object
(such as an ISM cloud, cf. \cite{fmn96}).
The contribution to the peak duration from
time the shell  takes to move through the cloud (i.e., $\Delta R_{\rm
amb}/[c\Gamma^2]$) is negligible compared to the time the shell takes 
to engage the perpendicular size of the object. 
This engagement time is caused by the curvature of the shell.
At an angle $\theta$
from the line of sight, the time to engage the object is 
$\Delta T_{\Delta R_{\perp}} = \theta\Delta R_{\rm amb}/(2c)$.  Note
that Table 2
of \cite{fmn96n} was incorrect for $\Delta
T_{\Delta R_{\perp}}$; see \cite{sp97a}.
At a typical angle of $\theta \sim \Gamma^{-1}$,
\eqnam{\RAMB}{RAMB}
$$ 
\Delta R_{\rm amb}  = {c\Delta T_p  \Gamma \over 2}
\eqno(\neweq)
$$
and
\eqnam{\SIZEB}{SIZEB}
$$
A_N = \pi \Delta R^2_{\rm amb} = {\pi c^2 \Gamma^2\Delta T_p^2 \over 4}~~.
\eqno(\neweq)
$$
Thus, both the case of shocks that grow from a seed and the case of 
shells running into
ambient objects are similar. If $c_s = c/3$,  then $A_N$ from equation
(\SIZEA) is 16/13 times larger than from equation (\SIZEB).
These two scenarios only differ  by a constant the order of unity.

The final ingredient for the calculation of the surface filling factor
is the area
of the shell visible to the observer:
\eqnam{\AREASHELL}{AREASHELL}
$$
A_S = 2\pi R^2(T)(1-\cos\theta_{\max})~,
\eqno(\neweq)
$$
where $\theta_{\max}$ is either the angular width of the shell [i.e., ${\rm
d}\Omega = 2\pi(1-\cos\theta_{\max}$)] or it is $\sim \Gamma^{-1}$, whichever
is smaller. Before the shell starts to
deaccelerate, presumably $\theta_{\max}$ is larger than
$\Gamma_0^{-1}$. Using $R(t) = 2\Gamma^2_0 c T$,
\eqnam{\AREACONST}{AREACONST}

$$
A_S = \pi \big[\Gamma_0^{-1}R(t)\big]^2 = 4\pi c^2\Gamma_0^2 T^2~~~~~~
{\rm if}~ \Gamma(T) = \Gamma_0~~.
\eqno(\neweq)
$$
Once the shell starts to deaccelerate,
\eqnam{\GAMMATIME}{GAMMATIME}
$$
\Gamma(T) = 463 \bigg[\INITIALCON\bigg]^{1/8} T^{-3/8}
 \eqno(\neweq a)
$$
$$
R(T) = 0.016~{\rm pc}~~ \bigg[\INITIALCON\bigg]^{1/4} T^{1/4} = ct
\eqno(\lasteq b)
$$
such that
\eqnam{\AREADEACC}{AREADEACC}
$$A_S = \cases {
  ~2.5 \times 10^{-4}~{\rm d}\Omega~{\rm pc}^2~
 \big[\INITIALCON\big]^{1/2}T^{1/2}
& if ${\rm d}\Omega < 2\pi(1-\cos\Gamma^{-1})$,  \cr
 ~3.67 \times 10^{-9}~{\rm pc}^2~
 \big[\INITIALCON\big]^{1/4} T^{5/4}
& if ${\rm d}\Omega > 2\pi(1-\cos\Gamma^{-1})$.  \cr
}  \eqno(\neweq)
$$
Combining equations (\EFF, \ENTITYRATE, \ENTITYNUM,
\AREACONST, \SIZEB, \GAMMATIME, \AREADEACC),
we find three different cases. Case a is the constant $\Gamma$ phase,
case b is the initial deacceleration when the size of 
the shell exceeds the beaming angle, and case c is 
when the deacceleration reduces the beaming such 
that  the shell's angular size  is no longer larger 
than the beaming angle.
For these cases, the filling factor is
\eqnam{\EFFEQ}{EFFEQ} $$ 
f = \cases{ N_N \big[{\Delta T_p \over T}\big]^2 {1 \over kf}=
   \big[{N^2 \over (\delta N)^2}-\sigma^2_{CS}\big] {\Delta T_p \over kfT}
& case a: $\Gamma(T) = \Gamma_0$,  \cr
N_N {\rm d}\Omega^{-1}~
10^{-6}~
\big[\INITIALCON\big]^{-1/4}
{\Delta T_p^2 \over kfT^{5/4}}
& case b: ${\rm d}\Omega < 2\pi(1-\cos\Gamma^{-1})$,  \cr
N_N ~{0.07 \over kf}~\big[{\Delta T_p \over T}\big]^2
& case c: ${\rm d}\Omega > 2\pi(1-\cos\Gamma^{-1})$,  \cr
} \eqno(\neweq)
$$
where $k$ is 16 for ambient objects and 13 for entities that grow from a
seed (see differences between equations (\SIZEA) and (\SIZEB).
\cite{sp97an} estimated the surface filling factor to be $\sim \Delta T/T$
(typically a few percent).
Our estimate of the filling factor approaches $\Delta T/(kT) \sim 0.1\Delta 
T/T$ when each peak is isolated
enough to be caused by a single entity. Thus, we will find smaller
surface filling factors than \cite{sp97a} for very spiky bursts and larger
values for smooth bursts.

The above was derived for the adiabatic case.  More generally,
equation (\GAMMATIME b) becomes
\eqnam{\GENGAMMATIME}{GENGAMMATIME}
$$
t \propto T^p = T^{{6a-(s+1+A)(2+a) \over s+A+7 }}~, \eqno(\neweq)
$$
where $A$ is 1 for the
adiabatic case and 0 for the radiative;  $s$ is a parameter that
varies from 0 to 1 depending on whether  most of the energy or
most of the mass is concentrated at low $\Gamma$'s; and $a$ is a
parameter that varies from 0 to 1 depending on whether the emitting
electrons are radiatively efficient or adiabatic (see \cite{rm98}).
More involved power laws are possible when one includes variations in
the ambient material (see \cite{mrw98}).
The surface filling factor from equation
(\EFFEQ) depends on ${\Delta T \over T}$. For a power law relationship
between $t$ and $T$, ${\Delta T \over T} = p^{-1}{\Delta t \over t}$.
Assuming that each entity is similar in the rest frame, different 
deaccelerations (i.e., various values of $A$, $a$, $s$) 
only modify the equation for  $f$ by
a numerical factor on the order of $p$.
We conclude that surface filling factors
found in the next section are roughly what one would find for any
deacceleration scenario as long as $t \propto T^p$.

\section{COMPARISONS TO OBSERVATIONS}

\subsection{Filling factor during the gamma-ray phase} \label{hairymath}

The consistency of the peak duration throughout the gamma-ray phase
implies that $\Gamma$ is rather constant.  In this case, the
surface filling factor is a function of $\eta$ (from Eq. [\EFF]), and the
direct observables ($N^2/\delta
N^2, \Delta T_p, \sigma^2_{CS}$, and $T$). 
The constant $k$ is either 16 or 13 depending on whether the source of
the entity is due to ambient objects or shocks that grow,
respectively. If ambient objects were involved, we would expect to see
peaks with a wide range of durations, perhaps a power law
distribution. In fact, most GRB peaks are about the same size 
(\cite{norris96}), so we will use $k = 13$ for the gamma-ray phase.
 To estimate $\eta$, one needs
to estimate the fraction of the visible shell (i.e., $A_S$, the area
within a beaming angle of $\Gamma^{-1}$) that contributes to the
signal during time $T$.  Unfortunately, we do not know how, in
general, to map the amount of active area of a shell to the observed time
history. Only in the particular case of a FRED-like burst can we
reliably estimate $\eta$.  For a FRED-like burst, the
curvature of the shell delays the photons from the region of the shell
at angle $\Gamma^{-1}$ by $T_0$ where $T_0$ is the time that the shell
expands before producing gamma-rays.  From \cite{fmn96n}
 and \cite{fs97n}, one can find $T_0$ from
the shape of the profile.  Given that the profile shape is
$(T/T_0)^{-\alpha-1}$, $T_0$ is $\sim 0.8 T_{50}$ where $T_{50}$ is the
duration of the burst that contains 50\% of the event counts (see
BATSE catalog, \cite{meegan96}).  Here, $\alpha$ is
the spectral number index, typically $\sim 1.5$. Thus, if one uses
$T_{\rm obs} = T_0 = 0.8T_{50}$, then the profile is from the region of the
shell out to an angle of $\Gamma^{-1}$, and $\eta \equiv 1$.

For long complex bursts, we cannot easily detect a FRED-like shape.
In lieu of an exact estimate, we will use the rule found for
FRED-like bursts: $T_{\rm obs} = 0.8T_{50}$. 
In Table 1, we list 6 FRED-like bursts and 46 long complex bursts that
were bright and had sufficient time structure for estimating
parameters. The first column gives the BATSE trigger number.  The
second column gives $\mu_N$ as found from equation (\ENTITYRATE).  
The third column gives an
estimate of $\Delta T_p$, the temporal width of a single entity.  In
most cases, this was estimated from the width of the autocorrelation
function.  For four of the FRED-like bursts, the profiles were rather
smooth (presumably due to a larger number of entities being
simultaneously active) so we used a typical value for $\Delta T_p$
(i.e., 0.5 s). The 4-th column is $T_{50}$ and the next two columns
are, respectively, the surface filling factor and 
its propagated error based on equation
(\EFFEQ).

In Figure 3 we show the distribution of surface filling factors
as a function of
burst duration $T_{50}$. The solid squares are the FRED-like bursts, and
the open squares are the long complex bursts.  Although some of the smooth
FRED-like bursts can have surface filling factors near unity,
most bursts have
values on the order of $5 \times 10^{-3}$.

\subsection{Filling factor during the x-ray afterglow}

In Figure 4, we show the time history of burst GB970828 observed by the
ASCA satellite, from 28.9 to 48.9 hours after the gamma-ray phase
(cf. \cite{yoshida98}).
At about 36.1 hours, there was a flare that lasted about 2
hours.
\cite{piro97n} also reported variations from a simple power law
decay in the x-ray afterglow of GR970508.
These two examples of deviations from a power law decay are quite
different.  In the case of the \cite{yoshida98n}  observation,
the flare is relatively
narrow (i.e., $\Delta T/T$ is small).  In the case of the \cite{piro97n}
observation, the deviation could be a step function increase in the
intensity
(i.e., $\Delta T/T \sim1$, although there is a data gap from 22 to 65
hr after the GRB that
obscures the nature of the deviation).
We will emphasize the \cite{yoshida98n} observation because the narrow
nature of the flare gives much stricter limits on the
relativistic shell (see \S\ 5).
Except for earth occultations, ASCA continuously observed the afterglow from
28.9 to 48.9 hr after the GRB.  The observations have been prorated into
samples of equal duration in Figure 4.

  Consider how case c in equation (\EFFEQ) applies to GB970828.
We must estimate the  fraction of the shell that
contributes to the 20 hours of observations, $\eta$.
We can estimate a lower limit on $\eta$, and therefore, an upper
limit on $f$.  In case c
we assume that the whole shell was active,
but only a fraction ($\eta$) of it contributed during the ASCA
observation.  
The narrowest annulus that contributes during any particular observation   
is one with a width equal to the size of a single entity.  
At a typical
angle of $\Gamma^{-1}$, $\eta$ can be as small as
\eqnam{\MINF}{MINF}
$$
{\rm min}~\eta = {\pi R_\perp \Delta R_\perp \over \pi R^2_\perp} =
{\Delta R_\perp \over R_\perp} = {\Delta T_p \over T}~~.
\eqno(\neweq)
$$
At smaller angles, $\eta$ would be smaller, but that is counteracted by
the fact that the peak in the ASCA observation is clearly much smaller
than the whole observation, such that the annulus responsible for the
observation must be larger than we used in equation (\MINF).

We only have
observations covering a small range of time, so we cannot use
equations (\ENTITYRATE) and (\ENTITYNUM) to estimate $N_N$.  
However, there are probably only a few active entities because
we see only one peak in Figure 4. Hence,
we will use $N_N =3$ as the maximum number of entities involved in the ASCA
flare. 
Thus, using case c, $f$ is, at most, $\sim 0.02(\Delta
T_p/T) = 10^{-3}$, a value very similar to what is obtained
in the gamma-ray phase.  In Figure 3, the ASCA flare for case c is
represented by the solid triangle.

In the situation of case b, we do not know \domega~ or $\eta$. However, the
smallest $\eta$ when the shell's angular size is limited to \domega~ is
\eqnam{\MINFB}{MINFB}
$$
{\rm min}~\eta = {\Delta R_\perp \over R_\perp} = {\Delta T_p \over
\Gamma T\sqrt{{\rm d}\Omega/\pi}}~,
\eqno(\neweq)
$$
such that 
\eqnam{\EFFMINB}{EFFMINB}
$$
f = {3.5 \times 10^{-8}N_N \over \sqrt{{\rm d}\Omega}}
 \bigg[\INITIALCON\bigg]^{-1/4} 
{\Gamma \Delta T_p \over T^{1/4}} =
{1.6 \times 10^{-5}N_N \over \sqrt{{\rm d}\Omega}}
 \bigg[\INITIALCON\bigg]^{-1/8} 
{\Delta T_p \over T^{5/8}}~~.
\eqno(\neweq)
$$
Using, $N_N = 3$ and $\INITIALCON = 1$, the maximum $f$ under case b
is $\sim 10^{-4}~{\rm d}\Omega^{-1/2}$.

In summary, $\Delta T/T$ in the ASCA flare is $\sim0.05$ which implies
a filling factor of $10^{-3}$.  The x-ray afterglow suffers from the
same time history problems that the gamma-ray phase has: peaks do not
have the value of $\Delta T/T$ expected from a shell whose surface
quasi-uniformly emits photons.

\section{DISCUSSION}

The rapid variations in GRB time histories imply emitting entities the
size of $\Delta R_\perp \sim c\Gamma\Delta T_p$.  Assuming a {\it
single} expanding shell, these entities must form on a much larger
surface, $\sim c\Gamma T$.  We have defined $f$ (given by
eq. [\EFFEQ]) to be the fraction of the surface of the shell which
becomes active, that is, generates emitting entities.  A crude
estimate of $f$ is $\sim 0.1 \Delta T_p/T$.  Because both the classic
gamma ray burst phase and the recently discovered x-ray afterglow
phase can have flares
with $\Delta T_p /T \sim 0.05$,
both time periods can suffer from an inefficient utilization of a single
relativistic shell.

In the case of the gamma-ray phase, there is a simple explanation: the
gamma-ray source is not a single relativistic shell, but rather,  reflects
activity at some central site that produces $\sim 10^{50}$ erg
s$^{-1}$ for up to several hundred seconds.  Other observations,
particularly gaps seen in the gamma-ray time histories
(\cite{frs98}), also 
argue for a central engine.  The x-ray afterglow, however, is widely
interpreted as being a single relativistic shell
(\cite{wrm97,danr97,wkf98,spn98,rm98}).
Therefore, we will
concentrate our discussion on how to accommodate flares during the
x-ray afterglow.  These arguments could also be applied to the
gamma-ray phase.

Of the two examples of deviations from a power law in the x-ray afterglow,
the \cite{yoshida98n} event (GB970828) is much more restrictive than the
\cite{piro97n} event (GB970508).  GB970828 has a narrow flare whereas
GB970508 appears to be roughly a step function. A step function is similar
to a FRED-like deviation which is approximately what one would {\it
expect} if most of the observable surface of a relativistic shell
interacts with a large ISM cloud (i.e., $\Delta T/T$ should be $\sim 0.25$,
\cite{fmn96}). Thus, GB970828 might be an afterglow with a low surface
filling factor whereas GB970508 has a high surface filling factor.

A common misconception is that one can just use an ISM cloud that covers
most of the shell's surface and, therefore, make a flare that
substantially changes the emission.  This does not work because the
curvature of the expanding shell prevents the shell from engaging the
cloud instantaneously.  Rather, the portion of the shell at $\theta \sim
\Gamma^{-1}$ requires a time $R(1-\cos\theta)/v$ longer to reach the
cloud.  Even if the cloud happens to have a concave shape such that the
shell reaches the cloud simultaneously over a wide range of angles, the
resulting photons at $\theta \sim \Gamma^{-1}$ must travel further to the
detector resulting in emission that is delayed by $R(1-\cos\theta)/c$. 
Indeed, the delay of the photons due to the curvature is identical to the
scenario of a shell that expands without producing photons for a long time
and then emits only over a small range of times (the $P=P_0\delta(t-t_0)$
case in \cite{fmn96}). The result is a FRED-like flare (like GB970508)
whose duration, $\Delta T$, is about $T$, rather than a small fraction of
$T$ as indicated by the ASCA observations for GB970828. 

The shell will have well-defined curvature because all points on it are
moving at nearly the same speed, that is, the speed of light. In any case,
there is no reason to believe that variations in the ambient material
would cause the shell to develop into a plane wave oriented towards the
observer such that the photons produced by an interaction with an ISM
cloud or a shock would arrive as a short flare. {\it Only the instantaneous
interaction between two plain parallel surfaces oriented perpendicularly
to our line of sight can produce a short peak from large surfaces.}

We see six possible explanations for flares with small $\Delta T/T$ that
produce large changes in the observed intensity. These explanations are
explained below and represented in Figure 5. A key issue is the size of
the corresponding energy reservoir required in each case. To intercompare
these energy requirements, we use that a typical peak corresponds to an
isotropic luminosity, $L_0$, equal to  $10^{51}$ erg s$^{-1}$. The total 
energy reservior is required to have
$$
E_{\rm tot} = L_0 {{\rm d}\Omega \over 4\pi} \epsilon^{-1} T_{\rm Dur}
\eqno(\neweq)
$$
where $T_{\rm Dur}$ is the average duration of the event (say, 30 s) and
$\epsilon$ is the fraction of the total energy that is converted into 
gamma-rays (say $10^{-1}$). If the emission is isotropic, then the total 
energy is about $E_{\rm iso} = 3 \times 10^{53}$ erg. Since we
cannot observe the energy released at angles
much greater than $\sim\Gamma^{-1}$, any model can reduce the required 
energy reservior to ${{\rm d}\Omega \over 4\pi}E_{\rm iso}$.  Although the
energy requirement can be made small,  a single shell
that emits nearly uniformly over \domega\ (see Fig. 5a) 
is not consistent with the observed time structure since a uniform
shell always results in a FRED-like shape. The following scenarios can 
produce bursts with peaks with small $\Delta T/T$:

\begin{enumerate}
\item{
The simplest way to have short powerful flares is to invoke a
central engine.  In a central engine, there are multiple releases of
energy at the central site over a time scale commensurate with the
observed duration of the GRB. Each release makes a peak in the time
history. Indeed, small $\Delta T/T$ and gaps are the key kinematic 
reasons why the GRB phase is likely to be
a central engine (\cite{fmn96,frs98}).
}
\item{
It is possible for  a large fraction of
the surface of the shell to be active, and the flare
region might still be small, because $\epsilon$ or the energy content of 
the shell  
varies strongly as a function of position on the shell.  That is, the 
photons cm$^{-2}$ s$^{-1}$ at the flare site must be
larger than the rest of the shell by a factor of $\sim f^{-1}$ (see Fig.
5b).  For example, some photon production rates scale as $\Gamma^3$, so a 
single region with a $\Gamma$ that is 10 times larger than other regions 
could 
produce emission that doubles the overall count rate. This would give 
the appearance that $f$ is $\sim 10^{-3}$. 
If the energy content of the shell is 
constant as a function of angle and $\epsilon$ varies, then the required 
total energy is about $f^{-1}{{\rm d}\Omega \over 4\pi}E_{\rm iso}$. If
$\epsilon$ is constant, and the energy content of the shell varies, then 
the total required energy is ${{\rm d}\Omega \over 4\pi}E_{\rm iso}$.
}
\item{
The third  explanation is that the local spherical
symmetry of the shell is broken into many emitting regions on a scale much 
smaller than $\Gamma^{-1}$ where only a 
fraction $f$ of the surface converts its energy into photons.
The regions are small enough ($\Gamma c \Delta T$) to make peaks with 
small $\Delta T/T$ (see Fig. 5c).
Here it is assumed that all regions of the shell has equal energy content 
but much of it is wasted because most regions do not produce gamma-rays. 
Thus, this scenario  requires an 
energy reservoir of $f^{-1}{{\rm d}\Omega \over 4\pi}E_{\rm iso}$, that is,
$f^{-1} \sim 10^3$ {\it more} energy than a simple shell. If \domega\ is 
large, this scenario could require $E_{\rm tot}$ to be as large as $10^{56}$
erg.  \cite{sp97an} invoke
this explanation and the resulting need for more energy to argue against a
single relativistic shell.  That arguement should not be considered 
particularly strong since we have several scenarios that do not waste the 
energy and/or \domega\ could be small.
}
\item{
The fourth explanation is that the local spherical symmetry is
broken because the outflow of material is in narrow fingers that
occupy only  a fraction $f$ of the shell's surface (see Fig. 5d).  This is
basically the same geometry as the previous case except those regions that
do not emit gamma-rays do not have energy content.
Since no energy is wasted in non-emitting regions, this 
scenario requires
an energy reservoir of only ${{\rm d}\Omega \over 4\pi}E_{\rm iso}$, that is,
about what one would expect from a simple shell.
}
\item
{ The fifth explanation is that \domega~is very small, so small that 
the emission lasts only $\Delta T_p$ even if the entire \domega~region is 
gamma-ray active.  This requires \domega~to be $\sim ({\Delta T_p \over
4T})^2$.  The requried total energy is correspondingly small: $({\Delta T_p
\over 4T})^2E_{\rm iso} = 10^{49}$ erg. This explanation has one narrow
ejecta that needs to
go through a gamma-ray emitting phase for each peak in the GRB time
history. In the above explanation 
(number 4), there are many narrow
ejecta, so each one only has to become gamma-ray active only a few times
to make a GRB time history with hundreds of peaks.
}
\item{The last explanation attributes the duration of the GRB to the
spreading out in time 
of the emission because of different  conditions as a function of
$\theta$ (see Fig. 5f). 
This is what we called a ``thick shell with substructure'' in 
\cite{fmn96}.
It would have been better to call it a ``thick emitting region with 
substructure'' since the actual emitting shell must always be thin (cf. 
\cite{fmn96}).
Effectively, $f$ is unity but it appears to be small because
the assumptions involved in equation (\AREASHELL) are not valid.
Let $t_i$  be the
time that the $i$-th section of a shell at $\theta_i$ expands before becoming
active. Let $\beta_i$ be the speed of the material at $\theta_i$.
The relative time that the emission from  two different regions ($i$ and $j$) 
arrive at our detector is
$$
T = T_i - T_j = t_i(1-\beta_i\cos\theta_i) - t_j(1-\beta_j\cos\theta_j)~~.
\eqno(\neweq)
$$
Explanations 3 and 4, above, assume that all $t_i$'s and $\beta_i$'s are
equal such that we can estimate the size of the shell from it's
FRED-like appearance and peaks arrive at different times because they
come from different $\theta_i$'s.
 Of course, most bursts are not FRED-like.
In this explanation, all $\theta_i$ regions become active (so $f$ =1),
but the peaks that would have added up to make
the FRED-like envelope are scattered
in time
because of the different
$t_i$'s and $\beta_i$'s.
To disrupt the FRED-like shape, $T_i$'s
must vary by  a factor of at least 2 from what they would be
if everything was constant with $\theta$.
 Consider the effect of different times of
emission (i.e., different $t_i$'s but similar $\beta_i$'s). The
$t_i$'s would have to vary by about a factor of 2 to change $T_i$'s by
a factor of 2.  Consider constant $t_i$'s but different $\beta_i$'s.
To spread out out the emitting regions so their distance from us varies 
by about $T$ requires that $(\beta_j-\beta_i)ct=T$ or that the radius of 
the shell is
$$
ct = {2\Gamma_i^2\Gamma_j^2 \over \Gamma_j^2-\Gamma_i^2}cT \sim 2\Gamma_i^2cT
\eqno(\neweq)
$$
The $\Gamma_i$ and $\Gamma_j$ would have to vary by about $\sqrt{2}$. Such
differences could arise from different baryon loading as a function of
$\theta$. There are two observations that argue against this
explanation. First, even the bursts with FRED-like envelopes
show low filling factors
(see the solid squares in Fig. 3). Second, if the $\Gamma_i$'s differ
by $\sqrt{2}$, then peak near the end of the burst should have widths
that are at least a factor of 2 wider.  Bursts normally do not 
appear to have  such a trend.  This explanation requires an energy 
reservoir like a uniform shell: ${{\rm d}\Omega \over 4\pi}E_{\rm iso}$.
}
\end{enumerate}
In conclusion, 
both the gamma-ray phase and the x-ray afterglow show similar rapid
variability, both have values of $f \sim 0.1\Delta T/T$ that
are similar. Such variability implies that either GRBs are central
engines, or that the local spherical symmetry is broken  on a
scale much smaller than $\Gamma^{-1}$.  In the last case, the
structure can either be due to variations in $\Gamma$ as a function of
$\theta$ (perhaps due to small scale variations in the baryon loading,
explanation 6 above), or due to only a fraction $f$ of the
shell becoming active (explanations 3, 4, 5 above).
The required total energy is $\sim f^{-1}E_{\rm iso}$, $E_{\rm iso},$ and
$({\Delta T_p \over 4T})^2E_{\rm iso}$ for 
explanations 3, 4, and 5, respectively.
Using typical values, the total energy can vary from $10^{56}$ to 
$10^{49}$ erg.  We note that explanation 5 (a jet much narrower than
$\Gamma^{-1}$) is the only explanation that can also easily explain gaps 
in the time history.

\acknowledgments  We thank Cheng Ho and Chuck Dermer for useful comments.  
Jay Norris provided the time histories analyzed for Figure 3.
 This
work was done under the auspices of the US Department of Energy and was
funded in part by the NASA GRO Guest Investigator program.

\clearpage
 
\begin{deluxetable}{rccccc}
\footnotesize
\tablecaption{GRB Surface Filling Factors. \label{tbl-1}}
\tablewidth{0pt}
\tablehead{
\colhead{Burst} & \colhead{$\mu_N$}   & \colhead{$\Delta T_{p}$ s}   &
\colhead{$T_{50}$ s}  & \colhead{$f$ \%} & \colhead{$\sigma_{f}$ \%}
} 
\startdata
     143&    1.5&   1.10&    6.7&   2.43&   0.812\nl
     219&    3.0&   0.51&    6.3&   2.37&   1.184\nl
     249&    1.8&   1.43&    6.7&   3.75&   1.312\nl
     257\tablenotemark{1}&   42.7&   0.50&   28.5&   7.21&   2.181\nl
     394&    2.8&   0.76&   37.0&   0.55&   0.157\nl
     467\tablenotemark{1}&    6.7&   0.50&    8.8&   3.65&   0.956\nl
     469&    5.2&   0.26&    4.0&   3.20&   0.450\nl
     563\tablenotemark{1}&   79.8&   0.50&    8.6&  44.41&   7.823\nl
     647&    4.7&   0.90&   16.8&   2.43&   1.636\nl
     660&    2.7&   0.26&    5.5&   1.22&   0.633\nl
     676&    2.8&   0.76&   18.4&   1.11&   0.668\nl
     678\tablenotemark{1}&    4.1&   0.32&   20.5&   0.62&   0.185\nl
     829&    3.4&   0.38&    6.0&   2.10&   1.740\nl
     841&    2.1&   0.45&    5.8&   1.54&   0.708\nl
    1122&    2.9&   0.51&    6.1&   2.38&   0.979\nl
    1141&   10.0&   1.54&    6.9&  21.47&   8.346\nl
    1288&    0.6&   0.89&  100.2&   0.05&   0.016\nl
    1440&    0.4&   0.51&    2.8&   0.69&   0.230\nl
\tablebreak
    1541&    1.0&   0.48&   10.6&   0.44&   0.255\nl
    1606&    2.9&   0.51&   53.4&   0.27&   0.088\nl
    1625&    3.7&   0.47&   10.0&   1.67&   0.557\nl
    1652&    2.6&   0.83&   23.9&   0.86&   0.468\nl
    1663&    3.9&   1.02&   10.6&   3.61&   1.370\nl
    1676&    1.3&   0.32&   30.8&   0.13&   0.032\nl
    1815&    2.5&   0.32&   10.8&   0.72&   0.249\nl
    1883\tablenotemark{1}&  120.2&   0.50&    3.6& 100&  13.420\nl
    2067&    2.4&   0.64&    6.9&   2.14&   0.918\nl
    2080&    4.0&   0.32&   22.7&   0.55&   0.328\nl
    2083&    2.4&   0.38&    7.6&   1.18&   0.319\nl
    2090&    0.3&   0.26&    8.6&   0.10&   0.039\nl
    2110&    3.6&   0.70&    7.5&   3.22&   0.475\nl
    2138&    0.9&   0.77&   12.0&   0.57&   0.214\nl
    2156&    1.1&   0.78&   92.0&   0.09&   0.031\nl
    2213&    0.3&   0.90&   51.6&   0.05&   0.013\nl
    2228&    1.5&   0.38&   30.3&   0.19&   0.038\nl
\tablebreak
    2329&    2.5&   0.70&    7.4&   2.32&   0.786\nl
    2436&    0.9&   0.90&   24.6&   0.32&   0.197\nl
    2450&    2.6&   0.26&   11.9&   0.54&   0.377\nl
    2533&    1.8&   0.45&   26.0&   0.30&   0.127\nl
    2586&    1.3&   0.26&    7.9&   0.40&   0.143\nl
    2700&    1.5&   0.32&   32.6&   0.14&   0.035\nl
    2798&    3.8&   1.02&    9.0&   4.19&   0.599\nl
    2812&    0.5&   0.38&   14.1&   0.14&   0.036\nl
    2831&    1.5&   0.51&   66.0&   0.11&   0.037\nl
    2855&    1.8&   0.83&   13.2&   1.11&   0.344\nl
    2889&    1.3&   0.38&   20.6&   0.24&   0.047\nl
    2929&    2.4&   0.19&   14.1&   0.31&   0.178\nl
    2984&    1.5&   0.26&   11.8&   0.31&   0.117\nl
    2993\tablenotemark{1}&    8.4&   0.26&   17.7&   1.17&   0.574\nl
    3035&    4.0&   0.58&   35.8&   0.62&   0.109\nl
    3057&    3.7&   0.77&   13.4&   2.05&   1.569\nl
    3128&    2.2&   0.26&   14.1&   0.39&   0.106\nl
  970828\tablenotemark{2}&&  7200&   $1.33\times 10^5$&   0.10&\nl
\tablenotetext{1}{FRED-like burst}
\tablenotetext{2}{Burst with flare in x-ray afterglow}
\enddata
\end{deluxetable}

%

\clearpage

%
\figcaption[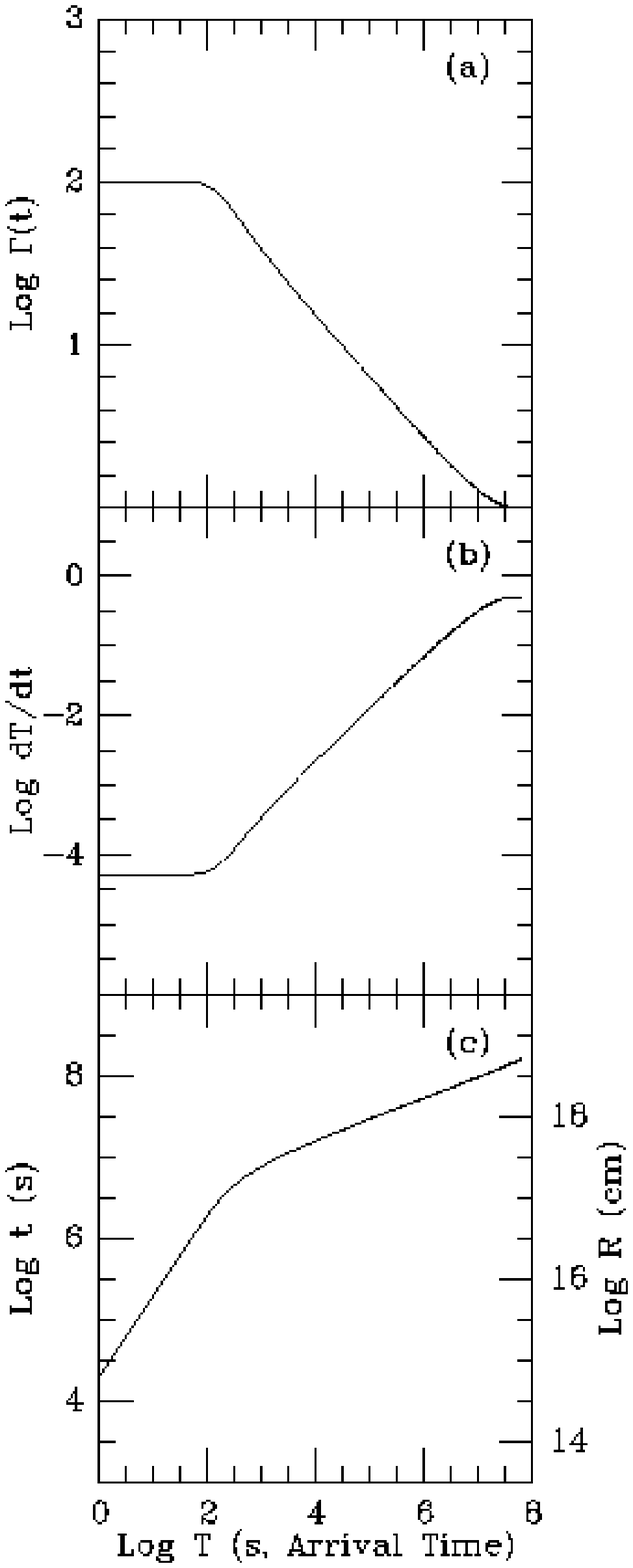]{
The effects of the relativistic motion on our perception of
GRB timescales.  The x axis is the time of arrival of photons at
a detector due to a relativistic shell moving directly at the detector.
This time is {\it not} the emission time in the detector's rest frame
which can
be measured only by clocks placed at rest with the detector at the location
of the emission.
\hfill\break
(a) The bulk Lorentz factor, $\Gamma$, as a function of the arrival time.
At first, the shell expands with a constant $\Gamma$.  Eventually, the
shell deaccelerates as it sweeps up the ISM, resulting in $\Gamma(T)$
varying as $T^{-3/8}$. Although we have  assumed only small energy losses
due to radiation, nonadiabatic solutions are similar.
\hfill\break
(b) The differential relationship between the time of emission as
 measured in the detector's rest frame time ($t$, the y axis) 
and the time of arrival of the
photons at the detector ($T$, the x axis). 
Because a shell moving directly at the detector keeps up with the photons it
produces, d$T$/d$t$ = $1-\beta(t) = \Gamma^{-2}(t)/2$.
 Note that no Lorentz transformations
are involved.  
\hfill\break
(c) The integral relationship between the detector's rest frame time and
the arrival time at the detector.  The left hand y axis is the time of 
emission in the detector's rest frame. The right-hand axis is the radius
of the shell at the time of emission (simply $c$ times the left-hand axis).
Note that during the gamma-ray phase, the
event appears to last for a short time (i.e., $T$ is a few hundred
seconds). In the detector's rest frame time, however, this corresponds
to a long time ($t \sim 4 \times 10^{6}$)
s.  The afterglow phase (where $t \propto T^{1/4}$) appears to cover
a wide dynamic range of time ($10^3$ to $10^7$ s).  Due to
deacceleration, it actually only 
covers a factor of 10 in time.
 \label{fig1}}

\figcaption[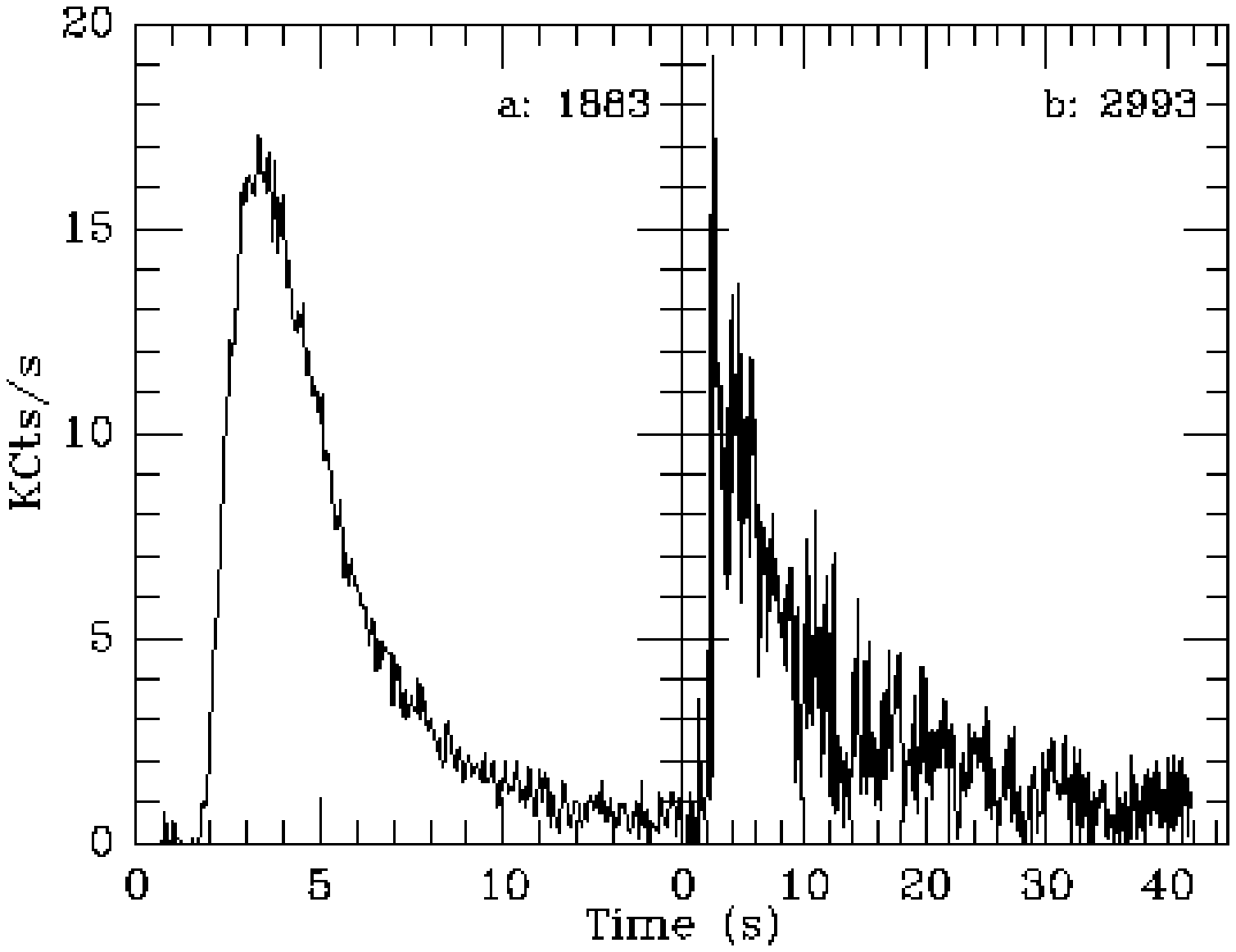]{
Two similar bursts with different surface filling factors, that is,
different fractions of the relativistic shell that become active.
A causally
connected region is much smaller than the available area of the shell.  The 
smoothness of BATSE burst 1883 implies that many casually connected
regions are simultaneously active, whereas the nonstatistical fluctuations
in BATSE burst 2993 implies that only a few regions are active. For burst
1883, we estimate that $1 \pm 0.13$ of the shell was active and for burst
2993, we estimate $0.12 \pm 0.006$.
\label{fig2}}

\figcaption[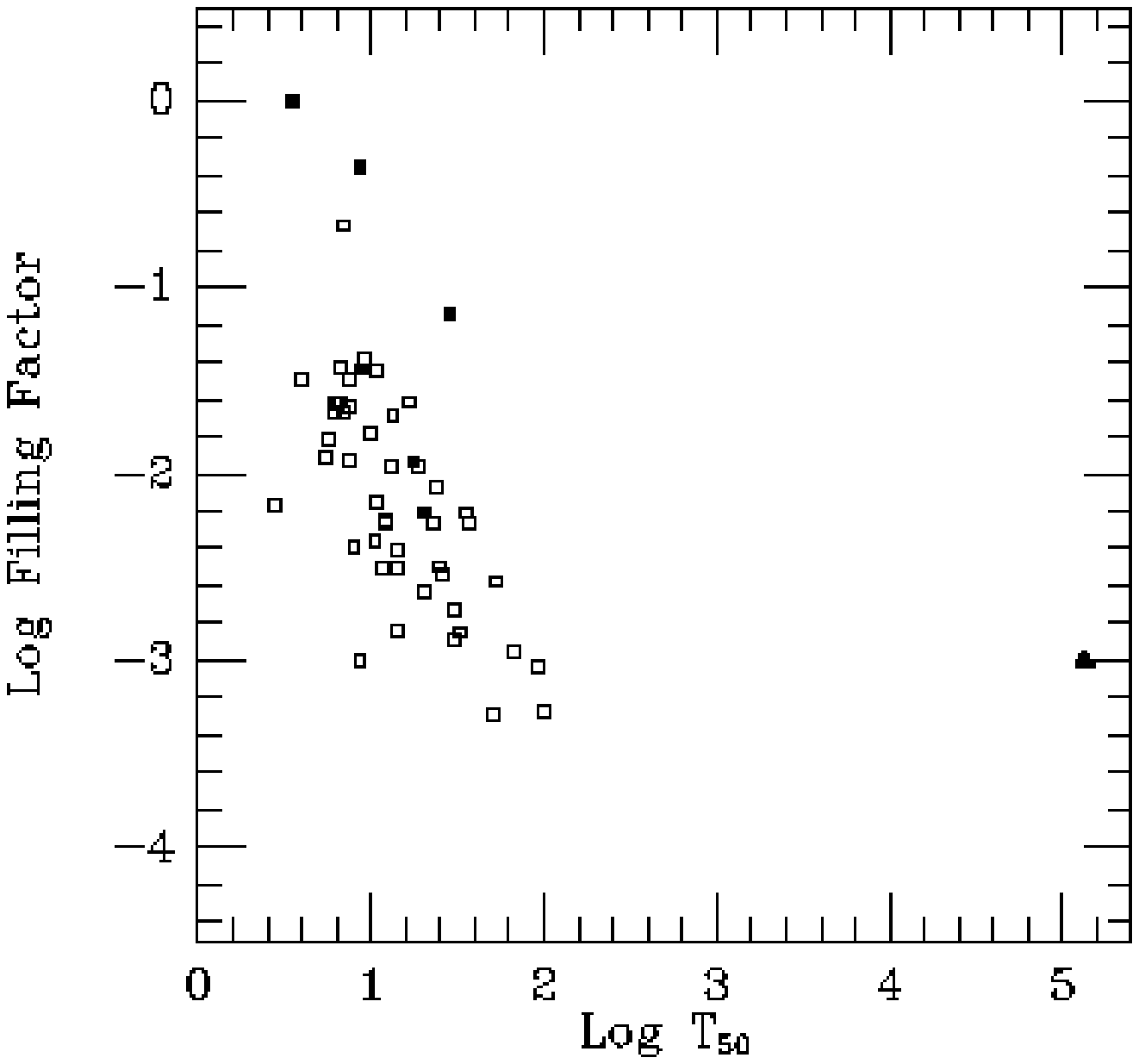]{
Typical values of the fraction of a relativistic shell that become
active during a GRB as a function of the duration of the emission ($T_{50}$
for the BATSE GRBs).  The six solid squares are FRED-like BATSE bursts for
which direct estimates of the size of the shell can be made.  The 46 open 
squares are long, complex BATSE bursts where we  estimated  the
size
of the shell  in a manner 
similar to the FRED-like estimates.
The single solid triangle is the surface filling factor
based on the 2-hr flare in the x-ray afterglow observed by 
ASCA at 36.1 hours after the
gamma-ray phase of GB970828.  Under most conditions, 
the surface filling factor is $\sim 0.1 \Delta T/T$.  Because $\Delta T/T$ is
similar for the gamma-ray phase and the
x-ray afterglow, they both give similarly low values for the
surface filling factor.
\label{fig3}}

\figcaption[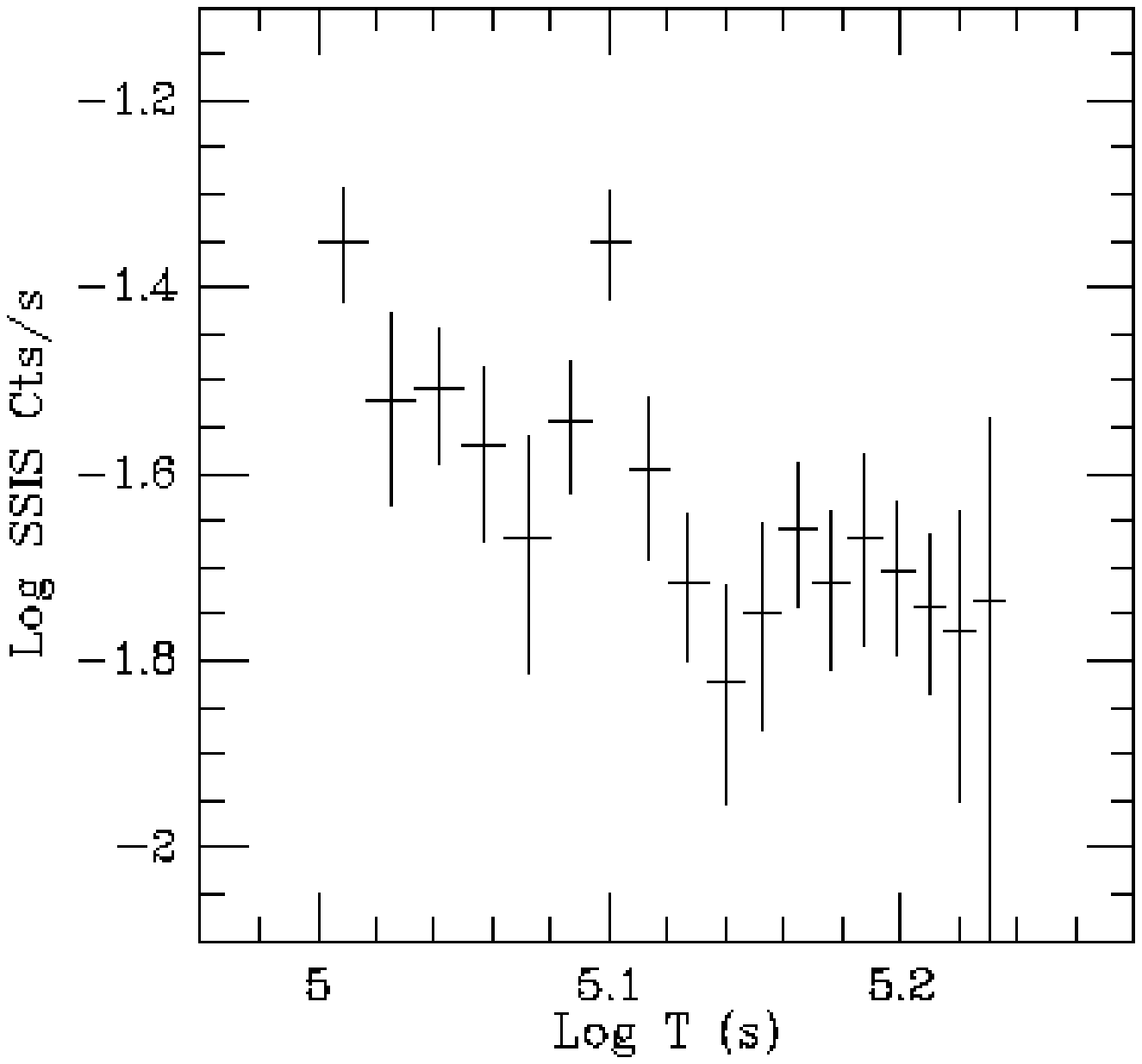]{
The x-ray afterglow of GB970828 as observed by ASCA.  At 36.1 hr
after the GRB, the x-ray afterglow had a flare that
lasted 2 hr which nearly
doubled the intensity.
The small value of $\Delta T/T$ for this flare ($\sim 0.05$) indicates that
the surface filling factor is small ($\sim 10^{-3}$ even during the
x-ray afterglow.
\label{fig4}}

\figcaption[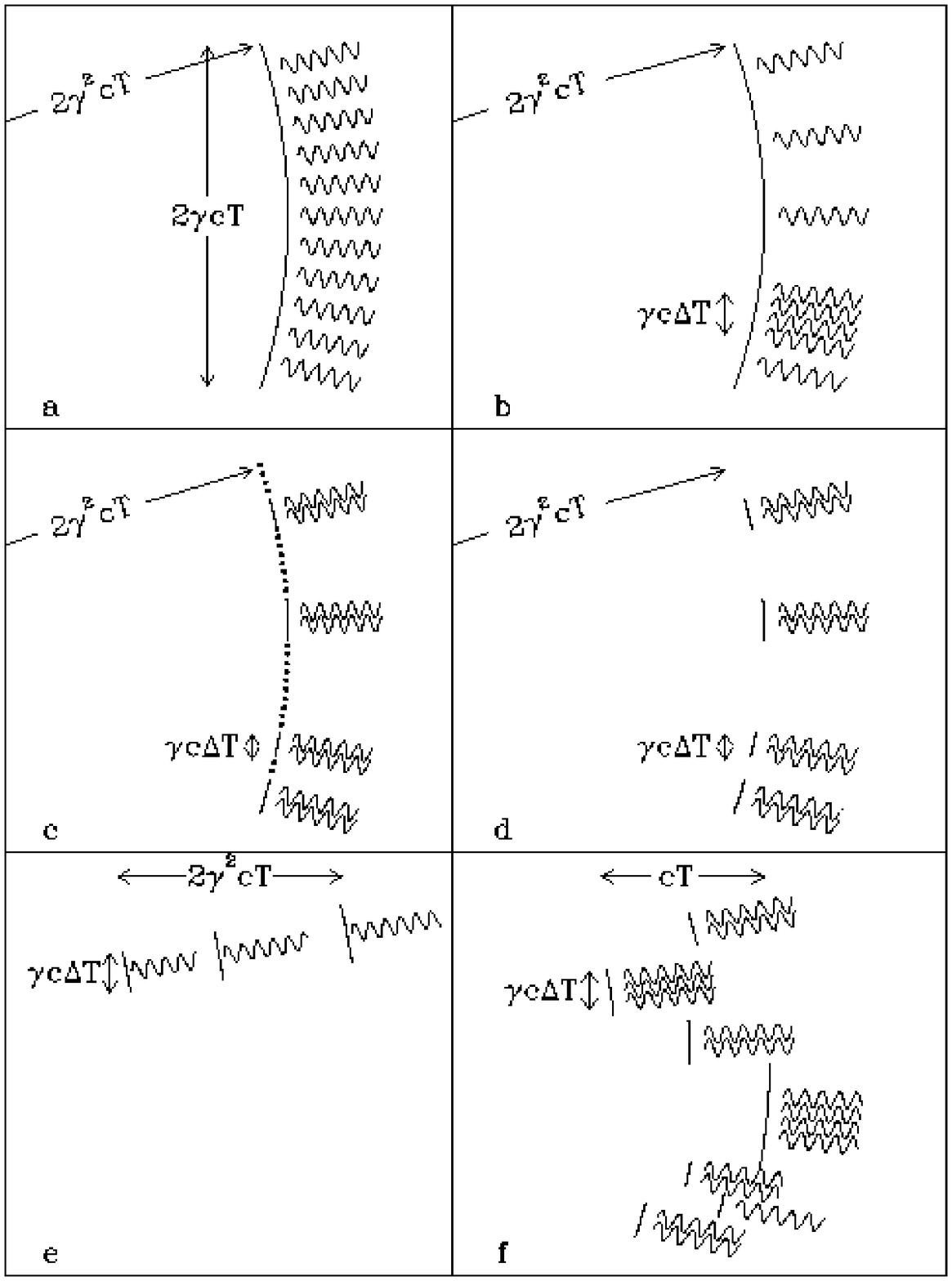]{
Source geometries related to producing the chaotic time 
histories that are typical in gamma-ray bursts.
\hfill\break
(a)A shell that emits uniformly over angles the order of $\Gamma^{-1}$ 
would have an emitting region with a transverse size of $\sim 2\Gamma c 
T$ and would produce a burst with a FRED-like time history.  The time 
history would be the sum of  many causally disconnected regions 
and, therefore, should be very smooth. The chaotic nature of GRBs 
indicate that the structure of the shell must be broken on a scale much 
smaller that $\Gamma^{-1}$. We characterize the chaotic time history by 
the ``filling factor'' ($f$) which is related to the maximum fraction of 
the causally disconnected regions that can be simultaneously active. A 
typical value of $f$ based on BATSE data is $\sim 5 \times 10^{-3}$.
\hfill\break
(b)Only structures the size of $\sim\Gamma c\Delta T$ can produce GRB 
with time structures of $\Delta T$.  One possibility is that a few small 
regions have a  much larger  efficiency for converting the bulk motion 
into gamma-rays.
\hfill\break
(c)Perhaps only a fraction $f$ of the shell becomes active.  In panel c, the 
dotted portions of the shell represents regions that never produce 
photons but has similar energy content. This explanation requires a 
factor of $f^{-1}$ more energy than the scenario of panel a  
since most of the bulk energy is never converted to 
gamma rays. This  could require so much energy that the 
concept of a single relativistic shell might have to be 
abandoned for the GRB phase. However, other scenarios do not required 
this much energy.
\hfill\break
(d)If the shell consists off many small, narrow jets and  the regions of 
the shell  between the jets are voids such that they do not waste bulk 
energy as in panel c, then the burst requires about the same energy 
than that of panel a. 
\hfill\break
(e)Perhaps \domega~is 
small enough, $\sim ({\Delta T_p \over 4T})^2$, that it can make time
structure 
the order of $\Delta T$. The required energy for this scenario can be as 
small as $10^{49}$ erg.
This is the only explanation that can also explain gaps in the time 
histories. \hfill\break
(f)Variations in the initial conditions as a function of $\theta$ could 
spread the causally disconnected regions out along the line of sight to 
the observer such that the resulting pulses do not make a FRED-like shape 
even though most of the shell within d$\Omega$ becomes active.
\label{fig5}}



\clearpage

\centerline{Figure 1}
\plotone{eff_asca_fig1.eps}
\clearpage

\centerline{Figure 2}
\plotone{eff_asca_fig2.eps}
\clearpage

\centerline{Figure 3}
\plotone{eff_asca_fig3.eps}
\clearpage

\centerline{Figure 4}
\plotone{eff_asca_fig4.eps}

\clearpage

\centerline{Figure 5}
\plotone{eff_asca_fig5.eps}

\end{document}